\documentclass{aa}

\usepackage{graphicx}
\usepackage{subfig}
\usepackage{txfonts,fmtcount}
\usepackage{natbib}

\usepackage[breaklinks=true]{hyperref} 
\bibpunct{(}{)}{;}{a}{}{,} 
\hypersetup{
    unicode=true,          
    colorlinks=true,       
    linkcolor=red,         
    citecolor=blue,        
    filecolor=magenta,     
    urlcolor=cyan          
}
\defcitealias{kronram}{K3}
\defcitealias{kapframmerge}{K8}
\defcitealias{dominik}{S1}
\defcitealias{kapfsluk}{K9}

\begin{document}

   \title{The collaborative effect of ram pressure and merging on star formation and stripping fraction}
   \author{%
      J.C. Bischko\inst{1}
      \and
      D. Steinhauser\inst{1}
      \and
      {S. Schindler\inst{1}}
      \offprints{Jan Christoph Bischko, \email{jan.bischko@uibk.ac.at}}
   }

   \institute{\inst{1}Institute of Astro- and Particle Physics, University
   of Innsbruck, Technikerstr. 25, A-6020 Innsbruck, Austria
   }

   \date{-/-}

   \titlerunning{Collaborative effect of ram pressure and merging}

  \abstract%
   {}
   {We investigate the effect of ram pressure stripping (RPS) on several simulations of merging pairs of gas-rich spiral galaxies. 
   We are concerned with the changes in stripping efficiency and the time evolution of the star formation rate. 
   Our goal is to provide an estimate of the combined effect of merging and RPS compared to the influence of the individual processes.}
   {We make use of the combined N-body/hydrodynamic code GADGET-2. 
   The code features a threshold-based statistical recipe for star formation, as well as radiative cooling and modeling of galactic winds. 
   In our simulations, we vary mass ratios between 1:4 and 1:8 in a binary merger.
   We sample different geometric configurations of the merging systems (edge-on and face-on mergers, different impact parameters). 
   Furthermore, we vary the properties of the intracluster medium (ICM) in rough steps: 
   The speed of the merging system relative to the ICM between $500$ and $1\,000$~km~s$^{-1}$, the ICM density between $10^{-29}$ and $10^{-27}$~g~cm$^{-3}$, and the ICM direction relative to the mergers' orbital plane. 
   Ram pressure is kept constant within a simulation time period, as is the ICM temperature of $10^7$~K. 
   Each simulation in the ICM is compared to simulations of the merger in vacuum and the non-merging galaxies with acting ram pressure.}
   {Averaged over the simulation time (1~Gyr) the merging pairs show a negligible 5\% enhancement in SFR, when compared to single galaxies under the same environmental conditions. 
   The SFRs peak at the time of the galaxies first fly-through. 
   There, our simulations show SFRs of up to $20$~M$_{\sun}$~yr$^{-1}$ (compared to $3$~M$_{\sun}$~yr$^{-1}$ of the non-merging galaxies in vacuum). 
   In the most extreme case, this constitutes a short-term ($<50$~Myr) SFR increase of 50~\% over the non-merging galaxies experiencing ram pressure. 
   The wake of merging galaxies in the ICM typically has a third to half the star mass seen in the non-merging galaxies and 5\% to 10 \% less gas mass.
   The joint effect of RPS and merging, according to our simulations, is not significantly different from pure ram pressure effects.}
   {}


\maketitle


\section{Introduction}
The changes galaxies undergo during their lifetimes are caused by complicated interplays between many physical processes (\citealt{lackgunn}). 
Isolating just a single contributor to galaxy evolution as the dominant one might not be possible (\citealt{mihosclustermerger}), and so the focus in theoretical work is generally shifting toward a more wholesome description of galaxy evolution. 
However, two physical processes, ram pressure stripping (RPS) and galaxy interaction, are thought to be the potent driving forces behind galaxy evolution.

Ever since \cite{gunngott} introduced the idea of RPS of studying galaxy evolution, this effect has received more and more attention: from early observations (e.g., \citealt{earlyramdwarf,cayatte}) to distant clusters and the existence of star formation in the stripped wake (e.g., \citealt{ramobserve4,hester}) that
\citet{abadi} and \citet{earlyram} pioneered in simulating RPS numerically, essentially confirming and refining the theoretical estimates of Gunn \& Gott.
In their simulations, \citet{vollmersim} and \citet{schulzstruck} found a mechanism for enhancing star formation in the inner region of galaxies not affected by RPS and discussed the role of RPS in the Butcher-Oemler effect (e.g., \citealt{butcheroehmler,butcheroehmler2}).
Simulations have been carried out using Lagrangian methods, for example, in order to quantify the RPS effects on internal kinematics \citep{kronkin}, star formation (\citealt{kronram},  \citealt{kapfsluk} hereafter \citetalias{kapfsluk}), and stellar bulge size (\citealt{dominik} hereafter \citetalias{dominik}).
Among others, \citet{roed06,roed07} used Eulerian grid codes, which complement the Lagrangian schemes.
\cite{tonsfr} and later \cite{roed14} explicitly target star formation in RPS tails.

Similarly, galaxy mergers are known to incite star formation activity (e.g., \citealt{mergerstarburst,interactstarearly}). 
The remarkable early simulations of \cite{toto} connected merger activity to the formation of tidal bridges and other features, leading to redistribution of gas to the core and thus star formation. The simulations of 
\cite{noguchi} show  that mergers could indeed trigger such an inflow. This general theory of merger-induced star formation has since been supported by a myriad of publications. 
More recently, \cite{hopstellarfeed,bournaudc} have focused on refining these models in an attempt to improve the resolution of individual giant molecular clouds as clumps of star formation. 
Additional effort has been put into modeling star formation processes correctly, starting with the \citeauthor*{kenni}-\citeauthor{schmidt} law
(see \citealt{schmidt,kenni}). Various improvements on these semi-empirical laws have been made since \citeauthor{schmidt}'s publication (see, e.g., \citealt{elmgreenstarformrev} for a general review). 

Owing to the high velocity dispersion in massive clusters, it has been long thought  that only high-speed fly throughs and fly bys would occur. 
Thus mergers, requiring low relative velocities between the merging partners, seemed very rare (e.g., \citealt{nomergeclust}). 
However, mergers do occur, because clusters form hierarchically and merge with infalling groups
of lower dispersion (\citealt{mihosclustermerger}). 
\cite{gnedin03} simulated that on average each galaxy has about one encounter with relative velocity $<500$~km~s$^{-1}$.
Depending on the criterion for cluster membership, the fraction of cluster galaxies that originate in groups varies (\citealt{luci12}).
\cite{mcgee09} estimate that, depending on cluster mass, between 30\% and 45\% of galaxies (M$_\star>10^9$~M$_\sun$~h$^{-1}$) are accreted from groups that are more massive than $10^{13}$~M$_\sun$~h$^{-1}$.
\cite{berri09} give a fraction of 10\% and find that $\sim$30\% of cluster galaxies are accreted with at least one luminous companion.

All this motivates a closer look at mergers in a cluster environment. 
To date, most theoretical work has regarded galaxy interactions and RPS independently. 
This approach is viable for disentangling the different phenomena, and it leads to an understanding of each phenomenon individually. 
In real galaxy clusters both processes occur at the same time. 
It cannot be said in advance whether both processes working together have synergetic effects, notably on star formation and stripping fraction. 
In this paper we expand the work of \cite{kapframmerge}, hereafter \citetalias{kapframmerge}, to a larger set of simulations and refine the techniques used there. 
That means that we explore a grid of merger and RPS parameters and make cross-comparisons to scenarios where one of the phenomena (merger or ram pressure) is missing. 


\section{Simulation}
\label{sec:sim}
We use the smoothed-particle hydrodynamics (SPH) and N-body code GADGET-2, developed by Volker Springel, for all simulations in this work.
Detailed information on the code can be found in \cite{gad}. The N-body part responsible for the gravity calculation uses a geometric \cite{gadgrav} oct-tree grouping with monopole moments in the force expansion and an error-limiting cell-opening criterion.
Long-range forces are calculated according to the TreePM method (e.g.. \cite{gadtreepm}) using a Fourier-based approach.
This is especially efficient for the periodic boundaries of our ICM setup (described below). GADGET-2 uses an energy- and entropy-conserving formulation of SPH (\citealt{Mon1,Lucy}) as presented in \cite{gadsph}.

\begin{table}
    \caption[Simulation parameters]{Important parameters used in all simulations (see \citealt{gadsfr,gad}).}
    \label{tab:simparas}
    \centering
    \begin{tabular}{rcl}
        \hline \hline
        \textbf{Parameter}          & \textbf{Symbol}   & \textbf{Value}    \\
        \hline
        Supernova fraction          & $\beta$           & $0.1$             \\
        Evaporation factor          & $A$               & $1\,000$          \\  
                Supernova temperature       & $T_\text{SN}$     & $10^8$~K          \\
                Cold cloud temperature      & $T_c$             & $1\,000$~K        \\
                Threshold density           & $\rho_\text{th}$  & $2.89 \times 10^{-25}$~g~cm$^{-3}$\\
                Star formation timescale    & $t_0^\star$       &  $2.1$~Gyr        \\
                Wind efficiency             & $\eta$            & 2                 \\
                Wind energy fraction        & $\chi$            & 0.5               \\
        \hline \hline
    \end{tabular}
\end{table}

\newcommand{\underterm}[2]{%
    \underset{%
        \text{%
            \textcolor[gray]{0.5}{#1}%
        }%
    }{%
        #2%
    }%
}

Implemented physics include radiative cooling (\citealt{gadradcool}) and, as described in \cite{gadsfr}, 
star formation, stellar feedback, and galactic winds. 
Relevant parameters of our simulations are summarized in Table \ref{tab:simparas}. 
Most notably Table \ref{tab:simparas} includes parameters for the statistical star formation recipe.
Star formation is implemented as a subgrid model, since single stars cannot be resolved individually.
Star
formation sets in when the density of a gas particle exceeds a threshold $\rho_\text{th}$.
The density of gas particles in star-forming regime is partitioned as a cold and a hot part, $ \rho_\text{gas} = \rho_c + \rho_h $.
Also considering the newly formed stars, the masses of these two phases evolve according to
$$ \frac{\mathrm{d}}{\mathrm{dt}} \rho_c = \underbrace{-\frac{\rho_c}{t_0^\star}}_{\text{SF}} -\ A\beta\frac{\rho_c}{t_0^\star} + \frac{1-f}{u_h-u_c}\Lambda_\text{net}(\rho_h,u_h) $$
$$ \frac{\mathrm{d}}{\mathrm{dt}} \rho_h = \underbrace{\beta\frac{\rho_c}{t_0^\star}}_{\text{SN gas}} + \underbrace{A\beta\frac{\rho_c}{t_0^\star}}_{\text{SN heating}} - \underbrace{\frac{1-f}{u_h-u_c}\Lambda_\text{net}(\rho_h,u_h)}_{\text{rad. cooling}} $$
New stars are formed out of cold gas on a characteristic timescale $t_0^\star$. It is assumed that massive stars ($M>\, \text{M}_\odot$, constituting a fraction $\beta=0.1$ in a Salpeter IMF) immediately blow up as supernovae with their gas added to the hot phase. 
Furthermore, cold gas is evaporated thanks to star formation with efficiency $A$ and radiative cooling transfers hot gas to the cold phase. 
In a similar way, energy exchange between the gas phases is modeled.
The implementation of this star formation recipe is, however, done in a statistical fashion to avoid the calculation of the mass exchange processes.
In the case of self-regulated star formation, the hot gas phase evolves toward an equilibrium temperature. 
This also determines the fraction in cold clouds and the star formation rate can be calculated as $\dot{M}_\star = (1-\beta) \frac{\rho_c}{\rho} \frac{m}{t_\star}$. 
Consequently, new star particles are spawned from a given gas particle with a probability proportional to $1-\exp\left[ - \frac{\dot{M}_\star \Delta t}{m} \right]$ in a time step of length $\Delta t$.
These particles conserve the phase space properties of the progenitor gas particle.
For more details, see \citet{gadsfr}.

In our simulations, we vary geometric properties and the mass ratio of the galaxy mergers, ICM densities and relative velocities to investigate the combined effects of RPS and minor mergers on SFR and stripping.
Overall, 20 core simulations with different combinations of parameters were conducted, capturing the first two encounters of the constituent galaxies and spanning a simulation time of $1$~Gyr each. 
Including simulations done for comparison purposes, the total number of different individual runs is 58. Table \ref{tab:simulat} summarizes our initial conditions.  
To have an acceptable number of simulations, we do not sample the whole grid of parameters. 
Rather we vary a standard case referred to as $wstd$ in only one parameter at a time. 
The parameters used in Table \ref{tab:simulat} are described in the following subchapters along with the galaxy and ICM model.

For the naming convention in Table \ref{tab:simulat}, 
the different simulations are named after their type and the main parameter(s) that deviate from the standard case labeled $wstd$. 
For instance, the name ``$w\rho27$'' is used for a simulation of an ICM density of $\rho_\text{ICM} = 10^{-27}$~g~cm$^{-3}$. 
Otherwise, $w\rho27$ is identical to the standard case in its parameters. 
Simulations starting with the letter ``$w$'' are conducted in our ``windtunnel'' setup described in Sect. \ref{sec:ICM}. 
These simulations are meant to model merging galaxies in ICM wind. 
The starting letter ``$i$'' indicates single (non-merging) galaxies in the ICM, which are necessary for later comparison. 
Mergers in vacuum have an ``$m$'' in the beginning of their designations. 
A simulation starting with none of these letters refers to a single galaxy in vacuum.

\subsection{Galaxy model}

\begin{table}
    \caption[Initial galaxy properties]{Initial galaxy properties for the largest galaxy in our simulations. Those of the smaller galaxies are printed in brackets (where different). Values are before evolution of $1$~Gyr, i.e., before self regulation. The first seven are free parameters.}
    \label{tab:galiniparams}
    \tabcolsep=0.1em\begin{tabular}{rclr}
        \hline \hline
        \textbf{Galaxy property}    & \textbf{symbol}& \multicolumn{2}{l}{\textbf{value (smaller galaxies)}}                                    \\
        \hline
        Halo concentr.              & $c$            & $7$                                                          &                           \\
        Circular velocity           & $v_\text{200}$ & $180$~km~s$^{-1}$                                            & \tiny{(113, 99, 90)}      \\
        Spin parameter              & $\lambda$      & $0.06$                                                       &                           \\
        Spin fraction               & $j_d$          & 0.02                                                         &                           \\
        Disk scale height           & $h_d$          & $0.1$                                                        &                           \\
        \hline
        \# gas particles            & $N_g$          & \multicolumn{2}{l}{300\,000 $\qquad$ \tiny{(75\,000, 50\,000, 37\,500)}}                 \\
        \# disk particles           & $N_d$          & $\frac{2}{3} N_g$                                            &                           \\
        \# halo particles           & $N_h$          & $\frac{4}{3} N_g$                                            &                           \\
        \hline
        Disk scale length           & $l$            & $6.52~kpc$                                                   & (\tiny{4.11, 3.59, 2.90}) \\
        Gas mass                    & $M_g$          & $4.06 \times 10^{\ 9}~\text{M}_\sun$                         & (\tiny{1.01, 0.68, 0.51}) \\
        Stellar mass                & $M_d$          & $2.71 \times 10^{10}~\text{M}_\sun$                          & (\tiny{0.68, 0.45, 0.34}) \\
        Total mass                  & $M_\text{tot}$ & $1.75 \times 10^{12}~\text{M}_\sun$                          & (\tiny{0.44, 0.29, 0.21}) \\
        \hline
        Gas resolution              &                & $ 1.36 \times 10^{4}~\frac{\text{M}_\sun}{\text{particle}}$  &                           \\
        Stellar resolution          &                & $ 1.36 \times 10^{5}~\frac{\text{M}_\sun}{\text{particle}}$  &                           \\
        Halo resolution             &                & $ 4.32 \times 10^{6}~\frac{\text{M}_\sun}{\text{particle}}$  &                           \\
        \hline \hline
    \end{tabular}
\end{table}

The galaxies are created by the initial conditions generator used in \cite{feedmerge}, kindly provided by Volker Springel. 
The theoretical groundwork for the generator was described in \citet{diskform}. 
The generator uses a radial \cite{herquistprofile} profile of scale length $a$ for the dark matter spatial density:
$$ \rho_{\text{dm}}(r) = \frac{M_{\text{dm}}}{2\pi} \frac{a}{r(r+a)^3}$$
It relates to the more common NFW (\citealt{NFW}) profile via $a= r_s\,\sqrt{2[\text{ln}(1+c)-c/(1+c)}$, where $c=r_{200}/r_s$ is the concentration parameter with $r_s$ the scale length of the NFW halo, and $r_{200}$ the radius enclosing a mean dark matter density of 200 times the critical density.
Gas and stars in the disk follow an exponential surface density profile:
$$ \Sigma_{*,g}(R) = \frac{M_{*/g}}{2\pi l^2} \mathrm{e}^{-R/l} $$
The profile falls with disk scale length $l$. 
This in turn is computed from the angular momentum as a function of the spin parameter $\lambda$, circular velocity $v_{200}$, and halo concentration $c$. 
In the case of gas particles, the vertical mass distribution is sampled from the overall potential using the equation of state.
In the case of star particles, it is modeled as an isothermal disk of relative disk height $h_d$. 
Additional parameters are gas and stellar disk mass fractions, $m_d$ and $m_g$, relative to total mass and disk mass, respectively. 
The parameters of the galaxies used here are summarized in Table \ref{tab:galiniparams}.
The parameters in Table \ref{tab:galiniparams} were mainly selected for their stability. 
In $4~Gyr$ of evolution, they showed no morphological peculiarities and yielded no unexpected results in terms of SFR.
We used pure disk galaxies, i.e., without a bulge component or HI-disk. 
The galaxies in our simulations differ only in their $v_{200}$ parameter, representing the circular velocity of the halo at $r=r_{200}$.
We varied only the mass of the smaller merging partner as seen in Table \ref{tab:galiniparams}.
This gives mass ratios of $\mu=\frac{1}{4}$, $\mu=\frac{1}{6}$ and $\mu=\frac{1}{8}$ compared to the largest galaxy in use. 
In all, the chosen parameters are fit to represent gas-rich spiral galaxies.
The baryonic mass of our largest galaxy constitutes 2\% of its total mass, where 15\% is gas.
The particle numbers of the different components are chosen empirically by the same method as used by \cite{kapfint}. 
There it was found that star formation and gas distribution are essentially independent of the gas mass resolution, as long as it is below $10^6 \frac{\text{M}_\sun}{\text{particle}}$. 
The collisionless disk and halo particles can be sampled at a higher mass per particle because they are numerically less problematic.

\begin{figure} \centering
    \subfloat[Initial gaseous disk]{\includegraphics[width=0.23\textwidth]{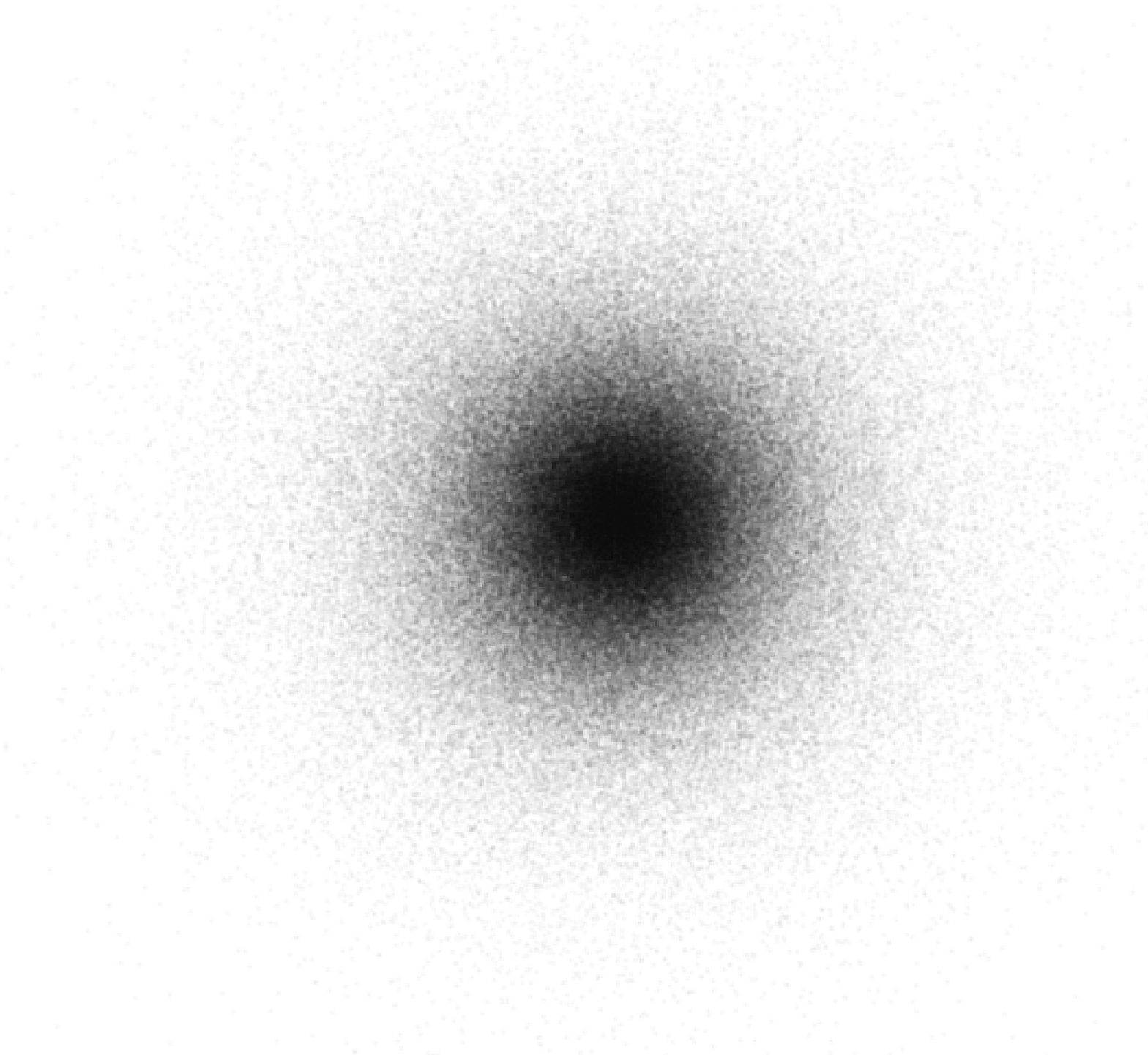}}
    \subfloat[Gaseous disk after $1~Gyr$]{\includegraphics[width=0.23\textwidth]{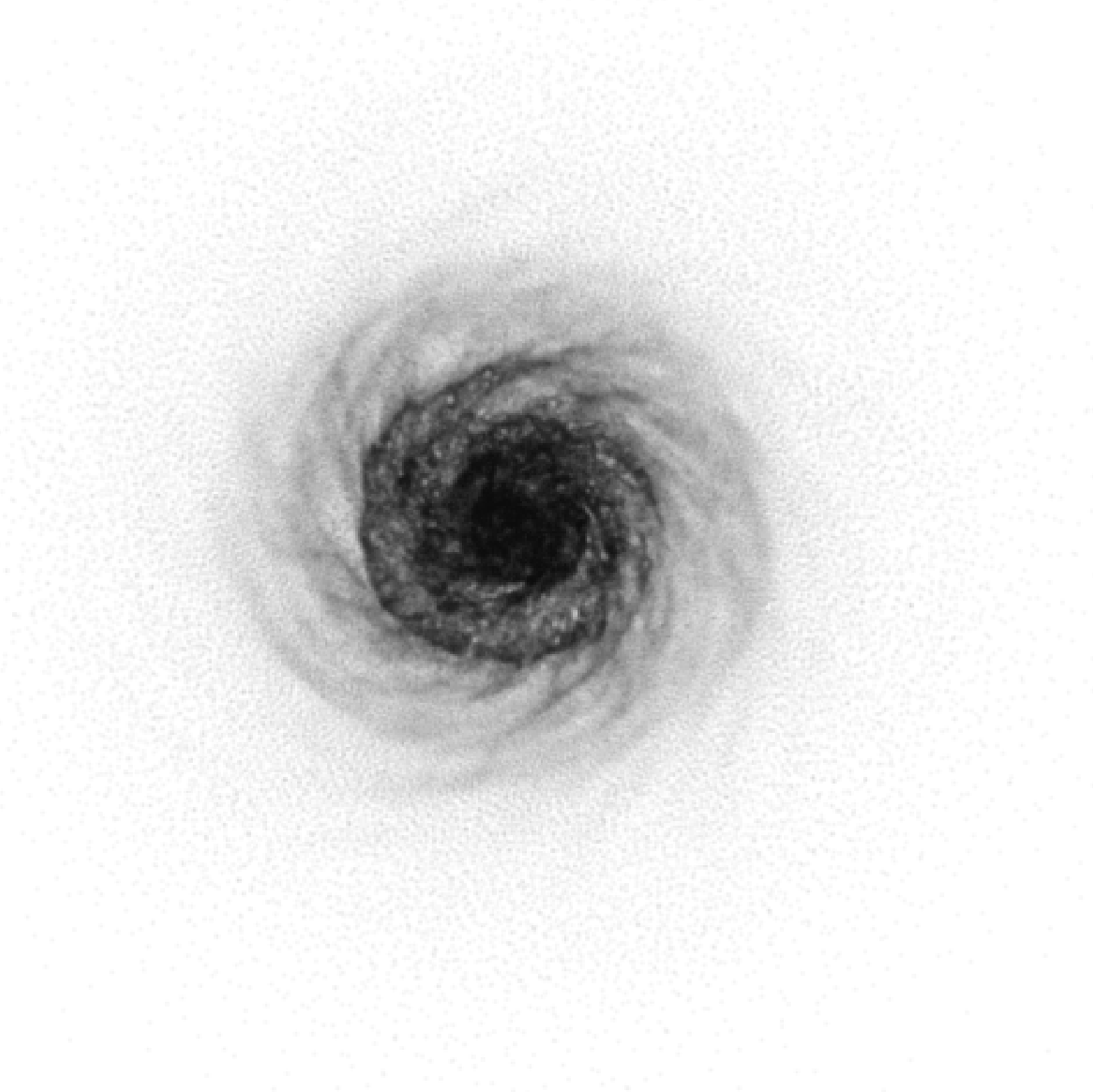}}
    \caption[Initial galaxies]{Particle distribution in disk galaxies created with the generator of \cite{feedmerge}. After $1~Gyr$ of isolated evolution spiral structure has developed and star formation self regulated (not shown).}
    \label{fig:inigal}
\end{figure}

The initial-condition generator provides a density and velocity distribution of particles. 
We evolve the created galaxies for $1$~Gyr in isolation. 
In this time, a spiral arm structure forms (see Fig. \ref{fig:inigal}) and $4.8\times10^{8}$~M$_\sun$ of stars are produced in the largest galaxy.
The radial density structure of our highest mass galaxy is very similar to the one described in \citetalias{dominik} (their Fig.1), but further extended owing to the higher circular velocity in our simulations.
After isolated evolution, the galaxies are paired up into binary mergers and, in case of the main runs, inserted into the ICM environment.

\subsection{ICM setup}
\label{sec:ICM}
GADGET-2 features only cubical simulation domains with periodic boundaries. 
For typical galaxy velocities within a cluster ($1\,000$~km~s$^{-1}$), this poses a computational problem: 
To capture at least the first two encounters of a merging pair of galaxies, the domain needs a size of about $L=1$~Mpc cubed. 
To simulate the ICM with GADGET-2, one could fill this domain with gas particles of sufficient mass resolution. 
To avoid numerical artifacts caused by the SPH scheme, the mass resolution of ICM particles has to match that of the galaxie's gas (\citealt{ottmassres}). 
Accurate simulations of both collisionless dynamics and smoothed particle hydrodynamics require a certain mass resolution. as does the star formation recipe. 
To fill the whole domain with particles of this mass resolution is computationally too expensive. 
We adapted the approach in \citetalias{kapfsluk} with a layered setup with resolution decreasing toward the outer layers and periodic boundaries.

\begin{figure} \centering
  \includegraphics[width=0.40\textwidth]{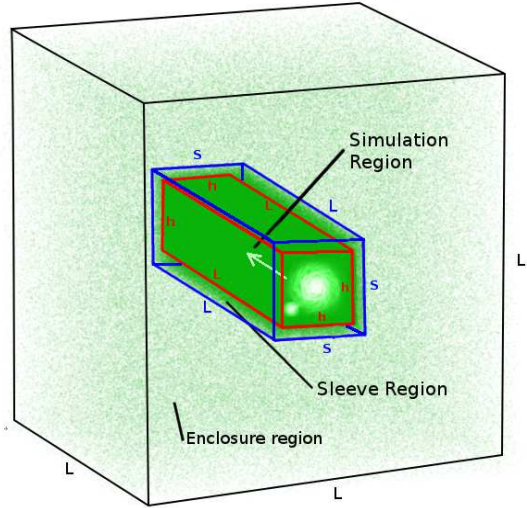}
  \caption[``Windtunnel'']{Three-layered ICM setup. Shown is the region of high resolution (inner border, $h\times h\times L$) in which the simulated merger takes place, the encompassing sleeve region of lower resolution (border $S\times S\times L$) and the rest of the simulation domain (outermost cube of edge length $L$). The particle's number density is indicated in shades of green. Galaxies move along the arrow inside the region of high-mass resolution.}
  \label{fig:windtunnel}
\end{figure}

Our setup consists of an inner elongated ``windtunnel'' of high-resolution particles encompassed by particles with lower resolved masses.
These coarser resolved particles keep the inner region stable. 
To prevent the setup from dissipating, we use periodic boundary conditions. 
In contrast to \citetalias{kapfsluk}, our setup introduces an additional layer, a sleeve region around the innermost part, which drops slightly in resolution. 
This improves stability of the inner region and allows for a steeper resolution drop in the large outer region, thus saving computation time. The setup is shown in Fig. \ref{fig:windtunnel}. 
The overall ICM density does not change by more than 2~\% in  the inner region of our simulations.
Dissipation and density fluctuations are suppressed better than in \citetalias{kapfsluk}. 
The mean distance of particles that have escaped the inner region from the region's edges is up to 20\% less in our setup than with \citetalias{kapfsluk}, while their number decreases by 30\%. 
Depending on the merger geometry, the ``windtunnel'' typically has a size of $120 \times 120 \times 1\,000 $ kpc with a sleeve width of $60$~kpc.
\footnote{Corresponds to $h\times h\times L$  in Fig. \ref{fig:windtunnel}. Sleeve width: $\frac{S-h}{2} = 60$~kpc.} This gives a time of $1$~Gyr for the galaxies to pass through.

The  high-resolution region is large enough to hold all of the galaxies' gas and most of the wake at all times during the merger. 
The cube size is sufficient for the whole extent of the galaxies' dark matter halo. 
It must be noted, however, that the periodic images of the galaxies still contribute to the overall potential. 
These are in principle unphysical and introduce small force errors. 
The ICM throughout this work is modeled with a temperature of $10^{7}$~K, which is consistent with typical ambient gas temperatures in a cluster.

In other ram pressure simulations, \citetalias{dominik} and \citet{jach07,jach09} find little dependence of SFR and stripping fraction on the direction of ICM flow with respect to disk orientation. 
Nearly edge-on cases (within $\approx 20^\circ$) only had significant deviations in stripping fraction from the cases where galaxies encountered the ICM with their disks oriented face-on to the ICM. 
However, the differences in SFR were negligible in all cases. 
Based on these results, and owing to other restrictions discussed in Sec. \ref{ssec:merger}, we chose to sample mostly face-on cases, with two edge-on ICM scenarios for completeness.
The ICM densities in our simulations range from $10^{-27}$ to $10^{-29}$~g~cm$^{-3}$.
Adopting a standard $\beta$ model\footnote{
    $\beta$-Model of \cite{beta2}: $\rho_\text{ICM}(r)=\rho_0 [ 1 + \left( r/r_c \right)^2 ]^{-3/2\beta}$.
    Parameters for the Coma cluster (\citealt{icm1}) are: $\beta = 0.7$, $r_c=0.26$~Mpc and $\rho_0=6.3\times 10^{-27}$~g~cm$^{-3}$.
}
for the ICM density, these values correspond to cluster-centric distances of 0.5, 0.8, 1.8, 2.6, and 5.6 Mpc for Coma, putting our standard case at 0.7 $r_\text{vir}^\text{Coma}$.
As such they cover typical densities from the outskirts to the regions where ram pressure dominates our simulations.
For Virgo our standard case would correspond to a distance of about 1 Mpc from the center and 0.3 Mpc for the highest density in our simulations (\citealt{sabinevirgo}, Fig. 11b).

The strength of ram pressure, as discussed by \cite{gunngott}, is given by
$$ P_{\text{ram}} = \rho_\text{ICM}\ v_\text{rel}^2 $$
where $P_{\text{ram}}$ is the pressure due to the relative movement of galaxies with respect to the ICM at speed $v_\text{rel}$. 
A gas density of the ICM, $\rho_\text{ICM} = \text{const.}$, is assumed in our simulations. 
The amount of stripped gas in this approximation is then determined by the radius at equilibrium between ram pressure and restoring gravitational force. 
A similar estimate for the stripping timescale can be made by comparing acceleration due to ram pressure with restoring acceleration due to gravity.

It was found, e.g., by \citetalias{kapfsluk} that SFR has a stronger dependence on $\rho_\text{ICM}$ than on $v_\text{rel}$.
Consequently, we sampled five different ICM densities in contrast to only two speeds within the ICM (see Table \ref{tab:simulat}).

\subsection{Merger setup}
\label{ssec:merger}

For the most general case of a binary merger, the centers of mass of the two galaxies approach each other at approximately Keplerian orbits. 
The angular momenta of their disks with respect to the orbital plane can have arbitrary orientations. 
The ICM flow may come from an arbitrary direction as well. 

However, the treatment of ICM in our simulations limits the range of some of the individual geometric parameters of the merger; that is, we can only simulate those geometries that fit within the high-resolution region of the ICM shown in Fig. \ref{fig:windtunnel}.
For convenience, we align the orbital plane with our ``windtunnel''. 
To capture the whole merging process from the start, the initial separation of both disks, $d$, must be at least $100~kpc$. 
The number is supported by both our own simulations and observations (\citealt{asmus1,asmus2,asmus3}). 
At this distance, gravitational interaction can be neglected for star formation and stripping, and the whole merging process is captured.

\begin{figure} \centering
  \includegraphics[width=0.48\textwidth]{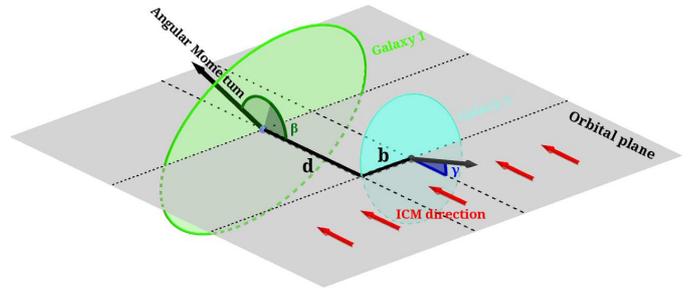}
  \caption[Simulation parameters (merger)]{%
    Geometric merger parameters as used in our simulations.%
    With $d$, the inital separation, $b$ the inital displacement, as well as $\beta$ and $\gamma$ the inclination of the larger and smaller galaxies with respect to the orbital plane.%
  }
  \label{fig:mergeo}
\end{figure}

\cite{asmus3} found significant enhancement in star formation only in galaxies at relative velocities $v \le 100$~km~s$^{-1}$, which is the highest relative initial velocity between two galaxies, which we chose to adapt in our simulations.
The effective set of geometrical parameters are illustrated in Fig. \ref{fig:mergeo}.

\newcommand{\retrangle}{ 0 {\tiny \textit{(180)}}}
\begin{sidewaystable*}
    \caption[Simulations and their designations]{Designation and properties of the different simulations conducted in this work.}
    \label{tab:simulat}
    \centering
    \begin{tabular}{l|cccccccccccccccc|r}
    \hline \hline
     \textbf{Parameter}    & \multicolumn{16}{c|}{\textbf{Simulation designation}} & \\
     \tiny{(where different)}               & $wstd(r)$\tablefootmark{a} & $w\mu4(r)$\tablefootmark{a}  & $w\mu8(r)$\tablefootmark{a} & $w\rho27$       & $w\rho528$          & $w\rho529$              & $w\rho29$    & $wvr500$   & $wvm50 $ & $wvm75$ & $wb10$ & $wb50$ & $w\gamma45$   & $w\gamma90$   & $wee$ & $w\beta\gamma$\tablefootmark{b}          & \textbf{Unit}\\
        \hline
        $\mu$              \tablefootmark{c} & 1:6           & 1:4           & 1:8       &   -             & -                   &   -                     &   -          &   -        &   -      &   -     &   -    &   -    &   -           &   -           &   -            &   -            & $1$ \\
        $\rho_\text{ICM}$  \tablefootmark{d} & $10^{-28}$    &   -           &   -       & $10^{-27}$      & $5 \times 10^{-28}$ & $5 \times 10^{-29}$     & $10^{-29}$   &   -        &   -      &   -     &   -    &   -    &   -           &   -           &   -            &   -            & g~cm$^{-3}$ \\
        $v_\text{rel}$     \tablefootmark{e} & 1000          &   -           &   -       &   -             & -                   &   -                     &              & 500        &   -      &   -     &   -    &   -    &   -           &   -           &   -            &   -            & km~s$^{-1}$ \\
        $v_\text{mer}$     \tablefootmark{f} & 100           &   -           &   -       &   -             & -                   &   -                     &   -          &   -        & 50       & 75      &   -    &   -    &   -           &   -           &   -            &   -            & km~s$^{-1}$ \\
        $b$                \tablefootmark{g} & 30            &   -           &   -       &   -             & -                   &   -                     &   -          &   -        &   -      &         & 10     & 50     &   -           &   -           &   -            &   -            & kpc \\
        $\gamma$           \tablefootmark{h} & \retrangle    & \retrangle    &\retrangle &   -             & -                   &   -                     &   -          &   -        &   -      &   -     &   -    &   -    & 45            & 90            & 90             & 90\tablefootmark{b} & $^\circ$ (deg.) \\
        $\beta$            \tablefootmark{h} & 0             &   -           &   -       &   -             & -                   &   -                     &   -          &   -        &   -      &   -     &   -    &   -    &   -           &   -           & 90             & 90\tablefootmark{b} & $^\circ$ (deg.) \\
        $d$                \tablefootmark{i} & 100           &   -           &   -       &   -             & -                   &   -                     &   -          &   -        &   -      &   -     &   -    &   -    &   -           &   -           &   -            &   -            & kpc \\
        \hline
        merger         
         & $m\mu6$  &$m\mu4$ &$m\mu8$
                                                     &   -             & -                   &   -                     &   -          &   -        & $mvm50$  &$mvm75$  & mb10   & mb50   & $m\gamma45$   & $m\gamma90$   & mee            & mee            & n.A.\\
        large galaxy      
        & $i\mu1$       & -       & -       & $i\mu1\rho27$   & $i\mu1\rho528$      & $i\mu1\rho529$          &$i\mu1\rho29$ &$i\mu1vi500$& -        & -       & -      & -      &  &   -           & -              &$i\mu1\beta90$  &  n.A.\\
        small galaxy\tablefootmark{j}       
        & $i\mu6$       & $i\mu4$ & $i\mu8$ & $i\mu6\rho27$   & $i\mu6\rho528$      & $i\mu6\rho529$          &$i\mu6\rho29$ &$i\mu6vi500$& -        & -       & -      & -      &$i\mu6\gamma45$&$i\mu6\gamma90$&$i\mu6\gamma90$ &$i\mu6\gamma90$ &  n.A.\\
    \hline \hline
    \end{tabular}
    \tablefoot{
      \tablefoottext{a}{Retrograde partner simulation, i.e., with $\gamma = 180^\circ$ end with ``r'', e.g. in $w\mu4r$ the two disk are rotating in opposite directions}
      \tablefoottext{b}{The galaxies in $w\beta\gamma$ merge edge-on, both with ICM face-on}
      \tablefoottext{c}{$\mu$, mass ratio between the two galaxies}
      \tablefoottext{d}{$\rho_\text{ICM}$, density of the ICM in the simulation domain}
      \tablefoottext{e}{$v_\text{rel}$, center of mass velocity of the merging galaxies relative to the ICM}
      \tablefoottext{f}{$v_\text{mer}$, initial velocity of the galaxies relative to each other}
      \tablefoottext{g}{$b$, impact parameter of the merging galaxies}
      \tablefoottext{h}{$\beta$ and $\gamma$, inclination angles to the orbital plane of the merger's more resp. less massive galaxy}
      \tablefoottext{i}{$d$, initial separation of the galaxies}
      \tablefoottext{l}{In the last three rows, the merger simulation in vacuum (starting letter ``m'') and the two single-galaxy simulations in ICM (``i'') are listed.}
    }
\end{sidewaystable*}


\section{Results}
\label{sec:res}

\subsection{Evolution of star formation rate}

\begin{figure}  
        \centering
    \scalebox{0.695}{ \input{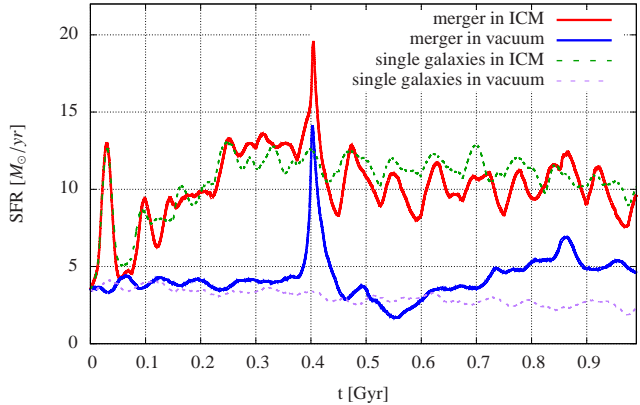}}
    \caption[SFR for $wstd$]{Overall SFR of the reference simulation $wstd$ and its comparison simulations, $i\mu6+i\mu1$ and $m\mu6$. Shown is the added SFR of the single galaxies in ICM and in vacuum (upper and lower dashed lines, respectively), as well as the SFR for the same galaxies merging in vacuum and in an ICM environment (solid lines). The strong peak in vacuum and ICM merger SFRs coincides with the galaxies first encounter at $t = 0.4 $~Gyr.}
    \label{fig:wstd}
\end{figure}

In this section we investigate the impact of the simulated ram pressure wind on our minor merger's star formation rate (SFR).
Figure \ref{fig:wstd} shows the evolution of SFR over $1$~Gyr of simulation time for our reference case, $wstd$, and its three comparison runs (non-merging galaxies in vacuum and ICM, merger in vacuum).
Similar plots are provided for the most interesting cases simulated. 
In the plots, we show the total SFR, including stars forming in the ram-pressure-induced wake behind the galaxies, as well as those in the galactic disks. 
In the next section (\ref{ssec:strip}), we discuss the distribution of newly formed stars between the two.

In most plots, the SFR shows two distinct peaks at about $25$ and $400$ Myr into the merger's evolution. 
The initial peak in the SFR plots (at $\approx25$~Myr) should be disregarded. 
It occurs when the galaxies first encounter the ICM of our setup. 
In reality they would gradually fall into a cluster's ICM as opposed to a sudden encounter, when inserted into the ``windtunnel'' (Sec. \ref{sec:ICM}). 
The disturbance caused by that, however, is overcome after the first $80$~Myr of evolution.
Simulations including ram pressure show oscillations in their SFR, which can be traced to density fluctuations in the galactic disk.
A possible cause is the numerical problem in SPH (see Sec. \ref{ssec:discuss}), seeing the compressions also occur when the star formation and galactic-wind subgrid model is disabled.
The oscillations, however, are not comparable to the physical effects that can be seen in our plots.

In Fig. \ref{fig:wstd}, the second and highest peak at a SFR of $19$~M$_\sun$~yr$^{-1}$ coincides with the time of the galaxies' first encounter ($t \approx 0.4 $~Gyr).
It is present in both curves that show the merger SFR in the ICM and the vacuum merger's SFR. 
As mentioned in the introduction, the origin of these peaks is mostly the triggered gas inflow to the center and higher gas densities due to the disturbance induced by the merging partner.

To put the mergers into perspective, the added vacuum star formation of the individual galaxies involved in the merger (i.e., added SFR of the two separate simulations $\mu1$ and $\mu6$) is also shown in Fig. \ref{fig:wstd}.
In a vacuum, the non-merging galaxies form stars at a slowly declining, combined rate of around $3$~M$_\sun yr^{-1}$. 
The vacuum merger has comparable SFRs up until the first encounter of the merging galaxies at $t=0.4 $~Gyr and then peaks sharply at $13$~M$_\sun yr^{-1}$. 
This increase surpasses the combined rate of the individual galaxies in the ICM without a merging partner (sum of SFRs for the simulations labeled ``$i\mu1$'' and ``$i\mu6$''), which is the top curve in Fig. \ref{fig:wstd}.
The first encounter peak in the ICM is 25\% higher than in the vacuum merger and higher than the non-merging galaxies in ICM.
This collaborative effect is a result of the core gas being more compact under ram pressure, making it more susceptible to gas inflows triggered by the merger.
The effect, however, is short-lived:
The second encounter peak at $t \approx 0.86$~Gyr is small in comparison to the vacuum case and barely visible in the ICM merger.
By the time of the second encounter, >25\% of the gas is stripped, and the gas and stellar disks of both galaxies are severely disturbed by the interaction.
These results are consistent with the results of \citetalias{kapframmerge}. 
We use more moderate parameters for the galaxies' gas fraction (15\% instead of 25\% in \citetalias{kapframmerge}), resulting in an overall lower SFR compared to \citetalias{kapframmerge} and \citetalias{kapfsluk}. 
For otherwise comparable galaxies, they have peak values of $26$~M$_\sun$~yr$^{-1}$ at an enhancement in ICM of a factor of $3.5$ to $8$.
The standard case represents the highest relative peak enhancement of SFR in our simulations.

\begin{figure}
    \centering
    \subfloat[ICM density $\rho_\text{ICM} = 5 \times 10^{-28} $~g~cm$^{-3}$\label{sfig:528}]{ \scalebox{0.68}{ \input{ wm528.tex    }} }\\
    \subfloat[ICM density $\rho_\text{ICM} = 5 \times 10^{-29} $~g~cm$^{-3}$\label{sfig:529}]{ \scalebox{0.68}{ \input{ wm529.tex    }} }
    \caption[Density ratio and SFR]{SFRs for different ICM densities. The higher density run $w\rho528$ in \subref{sfig:528} shows a minute merger peak and is otherwise indistinguishable from the single galaxies in ICM. Run $w\rho529$ in Fig. \subref{sfig:529}, on the other hand, has a visible first encounter peak in ICM that is slightly smaller than  in $wstd$ but that still surpasses the peak value in a vacuum.}
    \label{fig:mdensi}
\end{figure}

The impact of changes in the ICM density is shown in Fig. \ref{fig:mdensi}. 
We omit the plots for $w\rho27$ and $w\rho29$, i.e. highest and lowest simulated ICM densities.
They do not differ significantly from their comparison runs for ``ram pressure'' -- only ($i\mu1\rho27$ + $i\mu6\rho27$ in case of $w\rho27$) or vacuum ($m\mu6$) in the case of $w\rho29$ -- and thus show a limit for possible collaborative effects.

In both the ICM cases in Fig. \ref{sfig:528} (merging and non-merging), overall SFR is steeply declining for
$\rho_\text{ICM} = 5 \times 10^{-28}$~g~cm$^{-3}$.
By the time of the first encounter, when the first SFR peak appears in the vacuum merger, more than 50\% of the gas is stripped.
Thus, we barely see an enhancement in the ICM merger's peak SFR over the non-merging galaxies in ICM in Fig. \ref{sfig:528}.
The vacuum merger peak in SFR at the first encounter in Fig. \ref{sfig:528} is below the value for the non-merging galaxies in ICM.
In contrast, the lower ram pressure runs, $w\mu529$ and $wstd$, \text{show} a distinct first encounter peak in the ICM merger.
Indeed, it is a handy heuristic for the simulations in this work: if any encounter peak in the vacuum merger is at least comparable in height to the overall SFR in ICM single galaxies, the peak can also be seen in the ICM merger. 

Varying the relative speed between ICM and the galaxies ($wvr500$) showed only a decrease in overall star formation by about 15~\%, when compared to the standard case. 
Otherwise, $wvr500$ showed the same peak heights and general behavior as the standard case and is therefore not shown in detail.
This behavior is expected, e.g. from \citetalias{kapfsluk}, because the ICM density is the more crucial parameter for SFR in traditional SPH simulations than is the speed relative to the ICM, although the lower ram pressure via $v_\text{rel}$ does affect the stripping rate (see Sec. \ref{ssec:strip}).

\begin{figure}
    \centering
    \subfloat[Mass ratio $\mu = $ 1:4]{ \scalebox{0.68}{ \label{sfig:wm4} \input{ wm4.tex    }} } \\
    \subfloat[Mass ratio $\mu = $ 1:8]{ \scalebox{0.68}{ \label{sfig:wm8} \input{ wm8.tex    }} }
    \caption[Mass ratio and SFR]{SFRs for different mass ratios between the merging galaxies. Fig. \subref{sfig:wm4}, showing $w\mu4$, has a more pronounced second encounter peak in both ICM and vacuum compared to $wstd$ and more even so to $w\mu8$ in \subref{sfig:wm8}.}
    \label{fig:mratio}
\end{figure}

The SFRs for different mass ratios are shown in Fig. \ref{fig:mratio}. 
Since the partner is more massive and can trigger a stronger starburst, the simulation with mass ratio 1:4 (Fig. \ref{sfig:wm4}) has more pronounced merger peaks in vacuum than the reference case.
In the ICM merger, this translates into slightly higher merger peaks as well, but not a proportionate increase.
With $\mu = 1:8$ in Fig. \ref{sfig:wm8}, we seem to be at the limit mass ratio for the merger to show any changes in SFR in ICM environments.
In the counter-rotating runs, such SFR plots proved to be very similar to their co-rotating counterparts in Fig. \ref{fig:mratio}, with the small increase in SF expected from theory (stemming from additional instability). 
They were omitted for brevity, but included in the summarizing Fig. \ref{fig:gas}.

\begin{figure}
    \centering
    \scalebox{0.695}{ \input{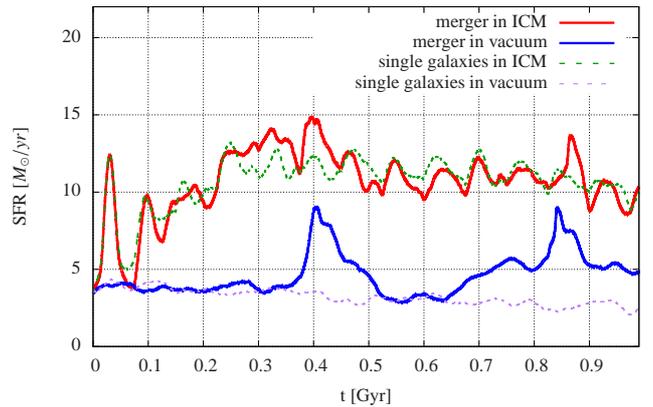}}
    \caption[Merger angle and SFR]{Simulation $w\gamma90$ and related SFRs with inclination angle of the smaller galaxy $\gamma = 90^\circ$. The galactic disks are oriented perpendicular to each other, resulting in a longer crossing time and thus broader but flatter vacuum merger peaks.}
    \label{fig:mangle}
\end{figure}

Other configurations of angular parameters, i.e., $\beta$, $\gamma$ (see Fig. \ref{fig:mergeo}), and the ICM direction are not all shown in distinct plots, but are part of Fig. \ref{fig:gas}.
Variations in these parameters flatten the merger-peaks in vacuum, because the interaction time between the two disk increases or the area for ram pressure to act on is reduced. 
Different displacements (impact parameters) and initial merger velocities, $b$ and $v_\text{mer}$, show little or no merger peaks in the ICM, since the interaction is too short. 
An example of this general trend is shown in Fig. \ref{fig:mangle} for $\gamma = $ 90$^\circ$. 
Here the smaller disk feels an edge-on wind, with the larger encountering the ICM face-on.
The vacuum merger peaks at around $t = 0.4~Gyr$ and $t=0.85~Gyr$ are broader compared to $wstd$ but less pronounced.
In ICM they are broader as well but barely stand out from the general enhancement due to ram pressure. 
Of these different geometric configurations, only $wb10$ has a sharp first encounter peak at $17$~M$_\sun$~yr$^{-1}$ ($15$~M$_\sun$~yr$^{-1}$ in the vacuum merger).
Just like $wstd$, the smaller galaxy hits the larger relatively close to its center in a face-on collision and thus yields a comparable outcome.
\begin{figure*}
    \centering
    \scalebox{0.9}{ \input{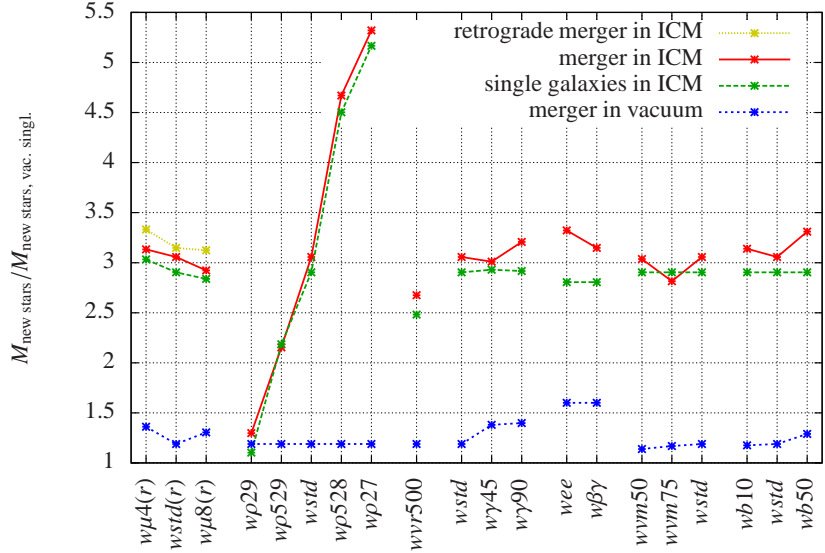} }
    \caption[]{Total mass of newly formed stars relative to the added total new star mass of the merger constituents in vacuum. Values are at $t= 1$~Gyr (end of simulation time). Mergers in the ICM show slightly but not significantly more star mass than the individual galaxies.}
    \label{fig:gas}
\end{figure*}

Finally, Fig. \ref{fig:gas} shows the relative mass of new stars at the end of the simulations relative to the corresponding mass in the individual galaxies in vacuum. 
\text{Only one }case ($wvm75$ with lower relative merger velocity) has formed noticeably fewer stars in the ICM merger than its constituent non-merging galaxies in ICM.
We attribute this slightly higher SF to stronger perturbations and triggered inflows to the star-forming center due to the merger, as well as the potential wells being deeper (added mass compared to single galaxies).
The variance in overall star formation with the individual merger and ICM parameters follows no clear trend and is comparably small. 
It ranges between almost no difference in the case of $\gamma=45^\circ$ to a difference in star mass of 18~\% between ICM single galaxies and ICM merger in the simulation with edge-on merger and edge-on ICM ($wee$).
The total mass fraction of new stars in the ICM mergers does not follow vacuum merger tendencies; e.g., relative mass of new stars increases slightly from $mvm50$ to $mvm75$ (vacuum mergers) but decreases between the ICM case ($wvm50$, $wvm75$).

With the exception of a narrow time frame around the first impact of the two merging galaxies, our simulations show negligible contributions of minor mergers to the SFR even in low-density environments (e.g., on the outskirts of typical clusters, outside the virial radius).

\subsection{Stripping and star formation in the wake}
\label{ssec:strip}

\begin{figure}
    \centering
    \scalebox{0.685}{ \input{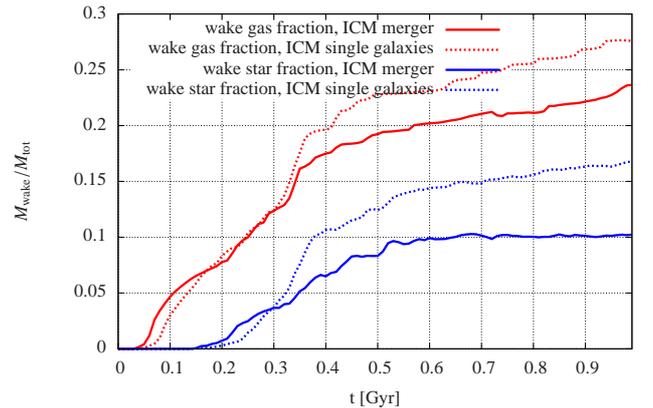}  }
    \caption[Stripping fraction for $wstd$]{Fraction of stripped gas to total gas mass of the galaxies (upper two lines) and mass of stars newly formed in the wake relative to the total mass of new stars (lower two lines). Shown are the values for both the merger (solid lines) and the added single galaxies (dotted lines) in ICM for $wstd$.}
    \label{fig:rstd}
\end{figure}

The time evolution of gas and star-mass fractions in the wake is plotted in Fig. \ref{fig:rstd} for the reference case $wstd$. 
We show these same fractions for the combined masses of the non-merging galaxies as well, in order to provide a comparison. 
With the star formation recipe used, wake stars form mainly in nodes of stripped gas (see \citetalias{dominik}, \citetalias{kapframmerge}).
Our criterion for wake stars and gas is simple: 
We consider particles to be part of the wake, if they lag $40$~kpc behind the common center of mass of the stellar disks. 
This way we can use a uniform criterion across our simulations, thereby covering all the unbound gas.
The result of a binding energy-based criterion for identifying wake gas agrees reasonably with this simpler criterion, with a delay of $\approx$50 Myr.

Initially all curves in Fig. \ref{fig:rstd} show a fast increase in wake mass fractions. 
When the gas stripping approaches the equilibrium predicted by \cite{gunngott}, the slope becomes shallower. 
In Fig. \ref{fig:rstd} this occurs after $t\approx 0.35$~Gyr for gas and $t\approx 0.5$~Gyr for the stellar component. 
In principle, the stellar component follows the trend set by the galaxies' gas with a temporal delay of just over $0.1$~Gyr. 
The increasing offset is a result of the reaccretion of wake stars by the galaxies, because stars no longer feel the ram pressure caused by the relative motion against the ICM. 
Notable as well is a difference in wake mass of both the stellar and gaseous components, when comparing single and merging galaxies. 
After the first $0.3$~Gyr, the single galaxies experience a higher stripping fraction than in the respective merger scenario. 
The added mass of the second galaxy in the merger, as well as the fact that the merging partner collects some of the stripped gas after first passage, explains this discrepancy.
The additional galaxy increases the potential barrier for gas to escape, once the galaxies are sufficiently close.
The companion galaxy overtakes the stripped gas in many of our simulations and binds some of  it again.

\begin{figure} 
    \centering
    \scalebox{0.685}{ \input{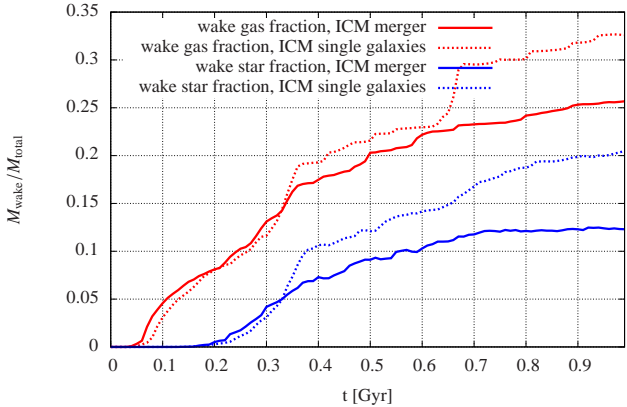}}
    \caption[Stripping for a lower mass fraction]{Stripping fraction at a merger mass ratio of $\mu = 1:8$. The curves shown here for the simulations $w\mu8$ in Table \ref{tab:simulat} are quantitatively the same as in Fig. \ref{fig:rstd} with slightly more gas being stripped.}
    \label{fig:rm8}

\end{figure}

In our mass regime and predominant merger geometry, this effect compensates for more turbulent gas and tidal features of the merger (such as the bridge), which are more readily stripped.
It does, however, depend on the exact merger geometry, as can be seen from $wee$ and $w\beta\gamma$, where re-accreation of gas hardly occurs.
Figure \ref{fig:rm8} shows the same plot for a smaller secondary galaxy. 
It is comparable to that of Fig. \ref{fig:rstd} with a slightly higher stripping fraction and higher ratio of star formation in the wake, although it shows a jump in stripping fraction for single galaxies at t$\approx$0.65~Gyr.
The cause of this jump is the core of the smaller galaxy finally being stripped. 
It is only weakly bound until that time.
In the ICM merger, the larger galaxy helps retain this core.

\begin{figure}
    \centering
    \subfloat[ICM density $\rho_\text{ICM} = 5 \times 10^{-28} $ gcm$^{-3}$]{ \label{sfig:r528} \scalebox{0.68}{ \input{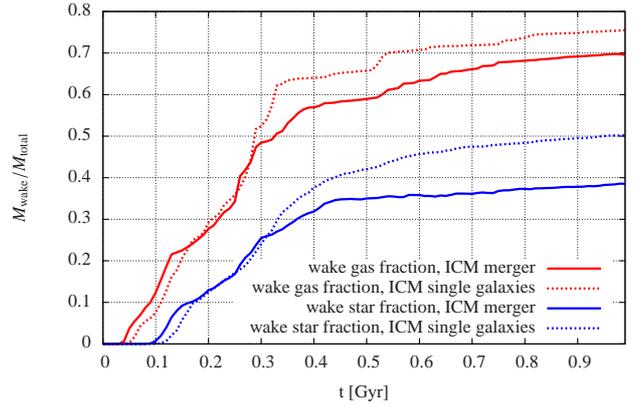}} }\\
    \subfloat[ICM density $\rho_\text{ICM} = 5 \times 10^{-29} $ gcm$^{-3}$]{ \label{sfig:r529} \scalebox{0.68}{ \input{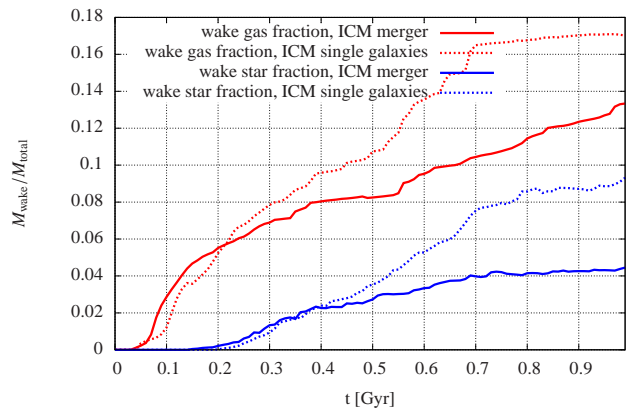}} }
    \caption[Stripping and density]{%
        Stripping fractions for the different ICM densities of the simulations $w\rho528$ and $w\rho529$. %
        Note the different scales on the ordinate. %
        The case for lower density \subref{sfig:r529} does not reach its saturation point within the simulation time of $1~Gyr$ and has a later onset in the wake star fraction.%
    }
    \label{fig:rdensi}

\end{figure}

Figure \ref{fig:rdensi} shows the situation at higher density (Fig. \ref{sfig:r528}) and lower density (Fig \ref{sfig:r529}).
In case of higher density (Fig. \ref{sfig:r528}), the stripping rate is enhanced by about a factor of three compared to $wstd$. 
About 40\% of stars are located in the wake by the time saturation is reached in the higher density ICM. 
The relative difference between the curves for single and merging galaxies becomes smaller at higher densities. 
At higher densities, the stripping is faster and more severe, giving the merger less time to influence the process.
Clearly, ram pressure is the dominant effect here.

At lower densities (Fig. \ref{sfig:r529}), gas stripping is a more continuous process, with no clear saturation phase. 
The instability caused by the merger enables gas to be stripped even at later times, when the single galaxies would have reached their saturation point. In the merger,
however, gas stripped at later times forms fewer stars than do single galaxies.
The disturbance in the gas disk is much greater then, allowing for less star-forming nodes in the wake.
At even lower densities (simulation $w\rho29$; plot omitted), only 5\% of gas is stripped by the end of the simulation time with less than 1~\% of all newly formed stars in the wake.

\begin{figure}
    \centering
    \scalebox{0.685}{ \input{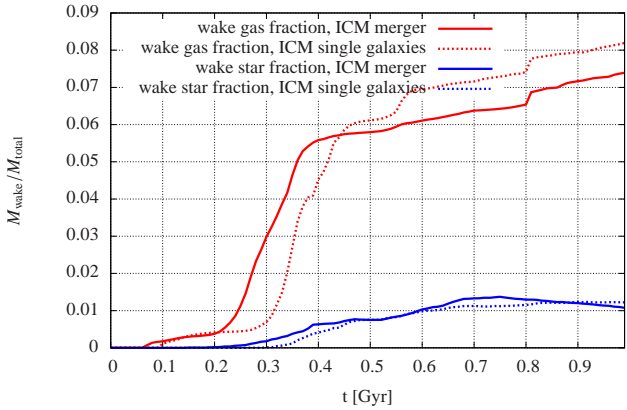}}
    \caption[Stripping fraction at lower relative velocity]{Stripping fraction for a lower relative velocity of $v_r = 500$~kms$^{-1}$ than in $wstd$ of Fig. \ref{fig:rstd}. The plot shows that significantly less mass is being stripped, and stars are noticeably re-accreted toward the end of simulation.}
    \label{fig:slwm6}
\end{figure}

In the graph of Fig. \ref{fig:slwm6}, where the stripping fractions of the runs with relative velocity of $v_r=500$~km$\,$s$^{-1}$ are shown, we see drastically decreased stripping and even less star mass in the wake.
Contrary to the situation for overall SFR in the previous section, a lower velocity relative to the ICM has a strong influence on the location of star formation (wake vs. disk). 
Toward the end of the simulation in Fig. \ref{fig:slwm6}, after $t=0.75~Gyr$, even re-accretion of stars is visible, as the lower solid curve declines slightly. 
This is the only simulation to show a decrease in the wake's stellar mass fraction.

\begin{figure*}
    \centering
        \subfloat[Gas fraction in the wake]         { \label{sfig:gasr}  \scalebox{0.68}{ \input{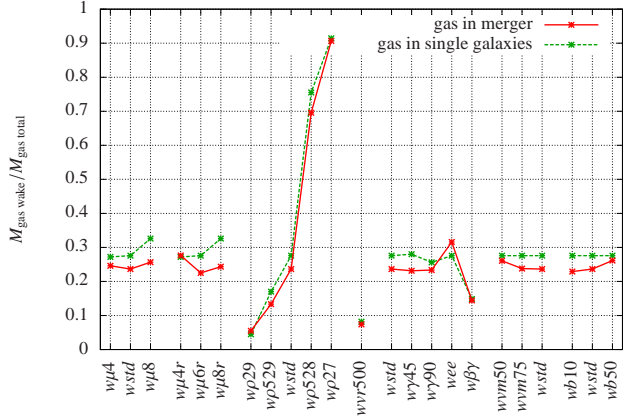} } }
        \subfloat[Fraction of new stars in the wake]{ \label{sfig:starr} \scalebox{0.68}{ \input{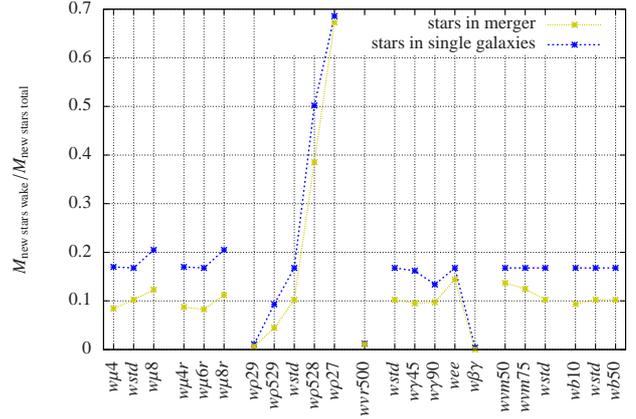} } }
        \caption[Gas and star fraction in the wake]{Gas and new star mass in the wake relative to the respective components' total mass ($t= 1 $~Gyr). Shown are the values for both the merger simulations in ICM and the added single galaxies in ICM that make up the respective merger.}
    \label{fig:ratio}
\end{figure*}

Figure \ref{fig:ratio} concludes this section by summarizing the results of all simulations. 
In Fig. \ref{sfig:gasr} one can see the amount of gas in the wake, again relative to the respective galaxies' total gas mass.
Shown are only the values at the end of the simulation. 
Geometric merger parameters show little effect on the amount of gaseous material in the wake between merging and non-merging galaxies.
Compared to the single galaxies in ICM, slightly less material gets stripped in the scenarios involving a minor merger, though this is dependent on merger geometry. 
The difference in stripped mass fraction between merging and non-merging galaxies in our simulations ranges from almost none ($w\beta\gamma$) up to $25\%$; for example,   $w\mu8$ has 26\% of its gas residing in the wake, while its constituent galaxies $i\mu8$ and $i\mu1$ combined have 33\%.

Figure \ref{sfig:starr} shows the same situation for the fraction of new stellar mass in the wake to total new stellar mass by the end of the simulations. 
Overall, the fraction of new stars formed in the wake is lower in the merger simulations compared to the single galaxies.
As little as half the fractional new stellar content in the wake of single galaxies can be seen in the wake of the respective merger cases.
This is to some extent a result of the added mass in the mergers.
More so, in most of our simulations the merging partner comes close to the wake of the other galaxy.
That enables the partner to accrete some of the stars from the wake of the other galaxy.
In the runs where the merging partner does not approach the wake ($wee$ and $wm\beta\gamma$) closely, the differences are smaller.
At higher densities, stripping is mostly complete before merger effects start to contribute. 
At lower densities, ram pressure stripping is no longer a factor, and the difference between merging and non-merging galaxies is small.
Only intermediate densities and favorable trajectories of the merging partner lead to collaborative effects.
In the light of our simulations, taking minor mergers into account in the process of ram pressure stripping does not yield significantly different results from ram-pressure-only scenarios (except for the wake's stellar content in the mentioned circumstances).

\subsection{Discussion and related work}
\label{ssec:discuss}
Unlike our simulations of constant ram pressure, real cluster galaxies experience varying $\rho_\text{ICM}$, depending on their path within the cluster. 
The recent study of \cite{bekki} uses a $\beta$-profile for $\rho_\text{ICM}$ and model ram pressure along different galaxy paths with an SPH code to quantify the effect of varying density on SFRs.
Their ``MW-type'' galaxy model is similar to the larger merging partner in our simulations with less total mass ($10^{12}$~M$_\sun$).
Their ``M33'' is most similar to our $i\mu8$.
The Virgo and group-like ICM models in \cite{bekki} yield a SFR enhancement by a factor of 2-3 (compared to our 3-5x enhancement), but only in the Gyr after the pericenter passage of the galaxies and very shortly before.
At this time in the galaxies' orbits within the Virgo model, ICM density would be $>5 \times 10^{-28}$~g~cm$^{-3}$.
This is different behavior from our individual galaxies in ICM that reach their enhancement sooner, after only 0.2 Gyr and generally show stronger enhancement.

A physical explanation may be that gradual infall removes gas gently without causing the conditions that increase SFR in our simulations.
With star-forming gas then missing at later times, enhancement in SFR at higher density might not be as drastic until the short-term peak density at pericenter passage comes into play.
It is difficult, though, to separate the physical effects from the different numerical models for star formation used in both our studies and other details of the simulation.
\cite{jach07,jach09} studied stripping rates using the same approach for varying $\rho_\text{ICM}$ as \cite{bekki}.
Their results indicate that short time variations in ram pressure do not affect the stripping rate.
Instead, the total encountered column density of ICM is the crucial parameter for the amount of material stripped.

We identified the deeper potential well and re-accretion by the merging partner's increased reach to be the cause of less pronounced gas stripping and fewer stars in the wake of galaxies merging in the ICM. 
We expect qualitatively similar stripping fractions under variable ram pressure.
When compared to simulations with adaptive mesh refinement (AMR), our simulations, and SPH simulations in general, differ greatly in SFR.
AMR simulations consistently show quenching rather than an increase in SFR under ram pressure. 
\cite{tonsfr} compare the approach of \citetalias{kapfsluk} to their AMR simulations and to observations. 
They attribute SPH' problems with resolving fluid instabilities and the resulting under-mixing, as well as the stiff equation of state as a cause for a high SFR in the tail. 
These and related problems with the standard pressure-entropy formulation of SPH have been discussed in \cite{codecomparis} and \cite{homix}.
The problems are most notable in the stripped tail, where much of the star formation takes place in dense nodes of gas.
This leads to increased stability and higher density of the star-forming nodes in the wake.
Generally in GADGET-2's formulation of SPH, strong density discontinuities lead to a false pressure estimate (which relies on a smooth density gradient), resulting in an unphysical separation force.
\cite{hesswindtunnel} showed that a gap between ICM and ISM can form because of that, which could cause the density oscillations in the disk.
This also keeps disk particles in the star-forming regime, especially at later times, when fluid instabilities and mixing would decrease star formation.
Observations support lower SFRs than in our simulations (\citealt{boss12}).

\cite{voll12b} finds possible enhancement factors of $1.5-2$ on the windward side of his sample galaxies in Virgo, while the overall effect is suppression of star formation, especially in the tail. 
Star-forming tails were found, among others, by \cite{ramobserve4} and more recently \cite{ramboserve6}.
The former estimates the total mass of current star bursts in the tail's nodes to be several $10^7$~M$_\sun$, asserting the total stellar mass formed in the tail could be a few times higher from already faded nodes.
\cite{ramboserve6}  shows strong galactic starbursts that are likely to be very short in duration. 
\cite{hester} investigated the Virgo dwarf galaxy IC3418, finding nodes of dense gas, generally rare in other Virgo galaxies, which \cite{jachnodes} studied in greater detail.
The total stellar mass in the tail of IC3418, however, is less than 1\% of the galaxies stellar mass with tail SFRs of $\sim 0.002$~M$_\sun$~yr$^{-1}$.
\cite{hesswindtunnel} compared the behavior of galaxies in a windtunnel between GADGET-2 and the moving mesh code AREPO (\citealt{arepo}). 
They demonstrated an overestimation by a factor of two in SFR and a smaller stripping fraction in SPH.\\
Our objective with this paper has been to determine whether minor mergers have an effect on stripping and SF that should be considered.
Although the issues discussed with SPH pose conceptional problems, we use relative quantities and compare, for instance, the relative increase in stripping fraction or SF between merging and non-merging galaxies in the ICM.
Moreover, our conclusion is that the effects of merging in the ICM are not significantly different from regular ram-pressure effects on single galaxies.

\cite{gad3merger} found that pure merger simulations using the standard density-entropy formulation of SPH compare well with results from state-of-the-art, grid-based hydrodynamic codes (i.e., AREPO).
The physical processes that could conceivably have led to synergies of ram pressure and merging (gas inflow to the galactic center, shocks, and disturbance of the gaseous disk from the merger, augmented by increased pressure from the ICM) are accurately modeled in the case of pure mergers or overestimated in SPH or while the counteracting stripping and mixing of the gas phases is underestimated.
Therefore the, in any case, small collaborative effects should be even less distinct in these codes, and our conclusion should hold.


\section{Conclusions}
\label{sec:con}
We have presented GADGET-2 simulations of merging galaxies in the ICM and compared them to galaxies evolved without a merger partner. Our simulations involved gas-rich spiral type galaxies in minor mergers (mass ratios between 1:4 and 1:8) and included their first two encounters.

In general, the merging galaxies retain 5\% to 10~\% more gas in their disk than do galaxies without a merger partner; i.e., less of their gas is located in the ram-pressure stripped wake. 
The added mass of the merging parter and re-accretion of wake gas by the partner (depending on trajectory) compensate for the disturbance of the gas disk that should make stripping easier.
Owing to re-accretion of stars and the added gravitational potential of the merging partners, we typically found 50\% fewer stars in the wake of merging galaxies.
This effect depends strongly on the trajectory of the galaxies with respect to the wind direction.
The relative differences are smaller in simulations with higher ICM density as ram pressure begins to dominate.

Even on the outskirts of clusters, with typical densities of $\rho_\text{ICM} = 10^{-28}$~g~cm$^{-3}$, our simulations indicate little overall effects on SFR owing to ongoing merging activity. 
We observe a short-term enhancement in SFR, compared to the merger in vacuum of up to 50~\%, but typically only 20~\%. 
These enhancements occur at the time of the first fly-through,  when ram pressure condensed the core gas, making it more susceptible to SF induced by the merging partner. 
Second encounters rarely show any SFR peaks in most cases.
Where they do, the enhancements are not higher than 20~\% over the added SFR of the individual galaxies in ICM. 
By the time of the second encounter, the galaxies' disks are too disturbed, and too much gas is stripped.
In any case the short-term enhancements do not last longer than $50$~Myr and do not contribute much to the overall star mass. 
When comparing the new star mass formed by the end of our simulations, generally the merging galaxies have formed slightly more stars, typically around 5~\% but up to 18~\% \text{more} than the individual constituent galaxies in ICM after a time of $1$~Gyr.
Depending mostly on ICM density and geometric parameters, the reason for this higher star formation in our simulations is a denser core and gas inflow triggered by the merging partner.
We discussed possible issues in Sect. \ref{ssec:discuss}.
They boil down to unphysical pressure estimates at the interface between ICM and ISM, possibly leading to an overestimation of the collaborative effects.
For overall star formation, the effect of varying geometric and ICM parameters related to ram pressure stripping dominates those of the minor merger parameters (mass ratio, merger geometry), and collaborative effects are small in any case.

\begin{acknowledgements}

The authors thank the anonymous referee, who helped to improve the quality of the paper significantly.
The authors acknowledge the UniInfrastrukturprogramm des BMWF Forschungsprojekt Konsortium Hochleistungsrechnen and the doctoral school - Computational Interdisciplinary Modelling FWF DK-plus (W1227).
DS acknowledges the research grant from the office of the vice rector for research of the University of Innsbruck (project DB: 194272).
The authors thank Volker Springel for providing the simulation code GADGET-2 and the initial-conditions generator.
We furthermore acknowledge profitable discussions with the colleagues at the institute.

\end{acknowledgements}

\bibliographystyle{aa}
\bibliography{collaborative_effect_final}

\end{document}